\documentclass[fleqn,usenatbib]{mnras}

\usepackage{newtxtext,newtxmath}

\usepackage[T1]{fontenc}

\DeclareRobustCommand{\VAN}[3]{#2}
\let\VANthebibliography\thebibliography
\def\thebibliography{\DeclareRobustCommand{\VAN}[3]{##3}\VANthebibliography}

\usepackage{graphicx}	
\usepackage{amsmath}	
\usepackage{mathtools}
\usepackage{hyperref}
\usepackage{xspace}	
\usepackage{euscript}

\usepackage{nccmath}

\usepackage[dvipsnames]{xcolor}	
\usepackage{pdflscape}
\usepackage{afterpage}
\usepackage{placeins}
\usepackage{bm}
\usepackage{ dsfont }
\usepackage{listings}
\usepackage{soul}
\usepackage{caption}
\usepackage{subcaption}
\usepackage{soul}
\usepackage{tabularx}

\usepackage{amsthm}

\usepackage{algorithm}
\usepackage[noend]{algpseudocode}

\definecolor{codegreen}{rgb}{0,0.6,0}
\definecolor{codegray}{rgb}{0.5,0.5,0.5}
\definecolor{codepurple}{rgb}{0.58,0,0.82}
\definecolor{backcolour}{rgb}{0.95,0.95,0.92}

\defcitealias{mwtrace2}{Paper~II}
\newcommand{\PaperII}{\citetalias{mwtrace2}}

\newcommand{\gaia}{{\it Gaia}\xspace}
\newcommand{\prob}{\mathrm{P}}

\definecolor{dkgreen}{rgb}{0,0.6,0}
\definecolor{gray}{rgb}{0.5,0.5,0.5}
\definecolor{mauve}{rgb}{0.58,0,0.82}
\title[The Milky Way Photo-Astrometric Tracer Density]{The Photo-Astrometric Vertical Tracer Density of the Milky Way I: \\ The Method.}

\author[A. Everall, N. W. Evans, V. Belokurov, D. Boubert and R. Grand]{
Andrew Everall,$^{1}$\thanks{E-mail: asfe2@cam.ac.uk}
N. Wyn Evans$^{1}$,
Vasily Belokurov$^{1}$,
Douglas Boubert$^{2,3}$,
Robert J. J. Grand$^{4,5,6}$
\\
$^{1}$Institute of Astronomy, University of Cambridge, Madingley Road, Cambridge CB3 0HA, UK\\
$^{2}$Magdalen College, University of Oxford, High Street, Oxford OX1 4AU, UK\\
$^{3}$Rudolf Peierls Centre for Theoretical Physics, Clarendon Laboratory, Parks Road, Oxford OX1 3PU, UK\\
$^{4}$Max Planck Institute for Astrophysics, Karl-Schwarzschild-Str. 1, Postfach 1317, D-85741 Garching, Germany\\
$^5$Instituto de Astrof\'isica de Canarias, Calle Vía L\'actea s/n, E-38205 La Laguna, Tenerife, Spain\\
$^6$Departamento de Astrof\'isica, Universidad de La Laguna, Av. del Astrof\'isico Francisco S\'anchez s/n, E-38206, La Laguna, Tenerife, Spain\\
}

\date{Accepted XXX. Received YYY; in original form ZZZ}

\pubyear{2021}

\begin{document}
\label{firstpage}
\pagerange{\pageref{firstpage}--\pageref{lastpage}}
\maketitle

\begin{abstract}
We introduce a method to infer the vertical distribution of stars in the Milky Way using a Poisson likelihood function, with a view to applying our method to the \textit{Gaia} catalogue. We show how to account for the sample selection function and for parallax measurement uncertainties. Our method is validated against a simulated sample drawn from a model with two exponential discs and a power-law halo profile. A mock \textit{Gaia} sample is generated using the \textit{Gaia} astrometry selection function, whilst realistic parallax uncertainties are drawn from the \textit{Gaia} Astrometric Spread Function. The model is fit to the mock in order to rediscover the input parameters used to generate the sample. We recover posterior distributions which accurately fit the input parameters to within statistical uncertainties, demonstrating the efficacy of our method. Using the GUMS synthetic Milky Way catalogue we find that our halo parameter fits can be heavily biased by our overly simplistic model, however, the fits to the thin and thick discs are not significantly impacted. We apply this method to \textit{Gaia} Early Data Release 3 in a companion paper where we also quantify the systematic uncertainties introduced by oversimplifications in our model. 
\end{abstract}



\section{Introduction}

Models for the distribution of stars in the Milky Way are key to stellar physics, Galactic archaeology (study of the formation history of the Galaxy) and understanding observations of external galaxies.

A core scientific aim of the \gaia mission is to map the 3D distribution of sources throughout the Milky Way \citep{Perryman2001}. To achieve this, \gaia has measured parallaxes for 1,467,744,818 sources \citep{Prusti2016, Brown2021} providing geometric distance estimates with no assumptions about source intrinsic brightness \citep{Lindegren2021ast}. However, we cannot straightforwardly use these billions of distances to construct a map of stars throughout the Galaxy for two key reasons. 

Until recently, the completeness limits of the \gaia catalogues were largely unknown. The observation strategy of the mission results in a completeness which varies significantly across the sky on sub-degree scales. Traditional methods of evaluating selection functions rely on the existence of a more complete source catalogue against which the sample can be compared; however there is no such catalogue to compare against, due to the incredible depth and resolution of \gaia across the entire sky. Without a selection function, it is impossible to generate an unbiased map of the Milky Way using the full power of the \gaia data. We refer the interested reader to \citet{Rix2021} for a detailed discussion on evaluating and using selection functions.

Furthermore, parallax-based distances are statistically awkward to work with. Much of our statistical methodology is constructed around the assumption of Gaussian measurement uncertainties, motivated by the central limit theorem. Parallax uncertainties are Gaussian distributed, which means that distances are reciprocal Gaussian distributed. This is a highly asymmetric distribution which, under an improper uniform prior, cannot be normalised. As such the distribution does not have a finite mean. Detailed discussions on how to use \gaia parallaxes for distance inference on individual stars are given in \citealt{BailerJones2015} and \citealt{Luri2018}.

In spite of these hurdles, the structure of the Milky Way has been studied in detail by many authors. A work-around to the challenges of parallax uncertainties is to focus on particular stellar populations for which the intrinsic brightness can be modelled. The distance can then be inferred from the measured apparent brightness. In some cases, simple stellar colour-absolute magnitude relations are used for either a large population of sources across the CMD \citep[e.g. ][]{Bilir2006sdss, Dobbie2020} or a small subset \citep[e.g. horizontal branch stars, ][]{Fukushima2019}. This approach has been taken further by using full stellar evolution models to infer intrinsic source brightness \citep{deJong2010}. Period-luminosity relations for certain variable sources are also incredibly valuable distance indicators. \citealt{Ak2008} used cataclysmic variables to estimate the vertical profile of the Milky Way disc, whilst \citealt{Mateu2018} used RR Lyrae to determine the structure of the old thick disc and radial profile of the halo. 

Some of these approaches apply uncertain colour-magnitude relations to large populations across the CMD, leaving the results susceptible to systematic biases. Other approaches use more carefully chosen sub-samples of specific stellar types such that only a small fraction of the data are used. 

In this paper, we develop a method to overcome the challenges of directly using \gaia parallaxes and applying the selection functions for the \gaia source catalogue and astrometry subset from \citealt{CoGV}. We demonstrate its feasibility on a \gaia-like mock sample with a known ground-truth. We limit the scope of this work to a high latitude region of the sky for statistical and computational reasons, and due to the challenge of dust extinction which we do not attempt to solve here.

The paper is arranged as follows. In Section~\ref{sec:method}, we introduce the likelihood optimization method used for this work, followed by a full description of the model in Section~\ref{sec:model}. The \gaia-like mock sample is explained in Section~\ref{sec:mock} and we demonstrate the application of the method in Section~\ref{sec:infer}. The method is tested on a more realistic mock catalogue in Section~\ref{sec:gums}. In a companion paper \citep[][henceforth \citetalias{mwtrace2}]{mwtrace2} we apply this method to high latitude regions of the \gaia EDR3 catalogue to estimate the vertical stellar profile of the Milky Way at the Solar radius and quantify the systematic uncertainties introduced by the simplifications and assumptions used in our model.

\section{Method}
\label{sec:method}

The probability of drawing a population of objects $\{\mathbf{x}_i\}$ from a density profile $\lambda(\mathbf{x})$ is given by the Poisson likelihood function (for which derivations are given in \citealt{Lombardi2013} and \citealt{seestar}),
\begin{equation}
\label{eq:poissonlike}
    \log\mathcal{L} = \sum_{i=1}^N \log\left(\lambda(\mathbf{x}_i)\right) - \int \mathrm{d}\mathbf{x} \lambda(\mathbf{x}).
\end{equation}
The observed population of objects is drawn from the true underlying distribution of sources multiplied by a selection function which gives the probability of a source being included in the survey. Therefore we can substitute $\lambda(\mathbf{x}) = f(\mathbf{x}, \boldsymbol{\psi})\mathcal{S}(\mathbf{x})$ where $\mathcal{S}$ is the selection function and $f$ is the true underlying source density with model parameters $\boldsymbol{\psi}$,
\begin{equation}
    \log\mathcal{L} = \sum_{i=1}^N \log\left(f(\mathbf{x}_i, \boldsymbol{\psi})\mathcal{S}(\mathbf{x}_i)\right) - \int \mathrm{d}\mathbf{x} f(\mathbf{x}, \boldsymbol{\psi})\mathcal{S}(\mathbf{x}).
\end{equation}
The aim of density estimation is to fit the parameters of the true underlying distribution, $\boldsymbol{\psi}$. Since the selection function is independent of the model parameters, it can be dropped out of the first term in the likelihood function, 
\begin{equation}
    \log\mathcal{L} \sim \sum_{i=1}^N \log\left(f(\mathbf{x}_i, \boldsymbol{\psi})\right) - \int \mathrm{d}\mathbf{x} f(\mathbf{x}, \boldsymbol{\psi})\mathcal{S}(\mathbf{x}).
    \label{eq:sflike}
\end{equation}
The source properties, $\mathbf{x}$, need to be chosen according to the dependencies of the model and selection function. 

\citealt{Mateu2018} use this method on a sample of RR Lyrae to constrain the structure of the thick disc and halo considering only spatial dimensions, while \citealt{Bovy2012} apply a more complex model to a population of G-dwarfs to fit the Milky Way disc using measured apparent magnitude, colour and metallicity. The aim of this work is to model the purely spatial distribution of sources, however the selection function, which will be introduced in more detail in Section~\ref{sec:sf}, is a function of position on the sky and apparent magnitude. Therefore, we must also consider the intrinsic brightness of a source, so our source properties are $\mathbf{x} = (l, b, s, M_G)$.

An additional complexity we introduce beyond previous works is accounting for parallax measurement uncertainties, which is vital when working with \gaia astrometry. Suppose instead that $\mathbf{x}$ are the \textit{measured} source properties and $f(\mathbf{x},\boldsymbol{\psi})$ is the expected distribution of measured source properties given the model. Source measurements are drawn from an uncertainty distribution, $\prob(\mathbf{x}\,|\,\mathbf{x}_{\rm T})$ where $\mathbf{x}_{\rm T}$ are the underlying \textit{true} source properties. The measured model ($f$) is given by a convolution between the true underlying model ($f_{\rm T}$) and the measurement error distribution,
\begin{equation}
    f(\mathbf{x},\boldsymbol{\psi}) = \int \mathrm{d}\mathbf{x}_{\rm T} \, \prob(\mathbf{x}\,|\,\mathbf{x}_{\rm T}) \, f_{\rm T}(\mathbf{x}_{\rm T}, \boldsymbol{\psi}).
\end{equation}
Substituting this into the likelihood, we get 
\begin{align}
    \log\mathcal{L} \sim \sum_{i=1}^N &\log\left(\int \mathrm{d}\mathbf{x}_{\rm T} \, \prob(\mathbf{x}_i\,|\,\mathbf{x}_{\rm T}) \, f_{\rm T}(\mathbf{x}_{\rm T}, \boldsymbol{\psi})\right) \nonumber \\
    &- \int \mathrm{d}\mathbf{x}_{\rm T}\, f_{\rm T}(\mathbf{x}_{\rm T}, \boldsymbol{\psi})\, \int \mathrm{d}\mathbf{x} \, \prob(\mathbf{x}\,|\,\mathbf{x}_{\rm T}) \,\mathcal{S}(\mathbf{x})
\end{align}
where we have reversed the order of integration in the second term and brought $f_{\rm T}$ outside the integral over measured parameters.

Our measured source properties are $\mathbf{x}=(l,b,G,\varpi)$, or Galactic longitude and latitude, apparent magnitude and parallax. In this work, we consider parallax error as the only significant measurement uncertainty. Positional uncertainties in $(l,b)$ are extremely small and we will test the impact of neglecting error in $G$ in \PaperII. Therefore, the error term becomes
\begin{equation}
    \prob(\mathbf{x}\,|\,\mathbf{x}_{\rm T}) = \delta(l-l_T)\,\delta(b-b_T)\,\delta\left(G-G_T(s,M_G)\right)\,\prob(\varpi\,|\,s).
\end{equation}
We integrate over all delta functions in the first term of the likelihood function
\begin{equation}
    \int \mathrm{d}\mathbf{x}_{\rm T} \, \prob(\mathbf{x}_i\,|\,\mathbf{x}_{\rm T}) \, f_{\rm T}(\mathbf{x}_{\rm T}, \boldsymbol{\psi}) = \int \mathrm{d}s \, \prob(\varpi_i\,|\,s) \, f_{\rm T}(l_i,b_i,G_i,s, \boldsymbol{\psi}).
\end{equation}

The selection function is a function of $l,b$ and $G$ only; there is no dependence on measured parallax \citep{CoGII, CoGV}. This makes it easy to integrate over
\begin{equation}
    \int \mathrm{d}\mathbf{x} \, \prob(\mathbf{x}\,|\,\mathbf{x}_{\rm T}) \,\mathcal{S}(l,b,G) = \mathcal{S}\left(l_T, b_T, G_T(s,M_G)\right).
\end{equation}
Finally, we can substitute this into the likelihood function,
\begin{align}
    \log\mathcal{L} &\sim \sum_{i=1}^N \log\left(\int \mathrm{d}s \, \prob(\varpi_i\,|\,s) \, f_{\rm T}(l_i,b_i,G_i,s, \boldsymbol{\psi})\right) \nonumber \\
    &- \int \mathrm{d}\mathbf{x}_{\rm T}\, f_{\rm T}(l_T,b_T,M_G,s, \boldsymbol{\psi})\,\mathcal{S}\left(l_T,b_T,G_T(s,M_G)\right).
    \label{eq:likelihood}
\end{align}
This is the likelihood function which we use to fit the model parameters, $\boldsymbol{\psi}$, to the observed data.
For the remainder of the paper, we will drop the subscript $T$ with $f$ always referring to the underlying source distribution.

\subsection{Parallax error integration}
\label{sec:p_integrate}

The biggest numerical challenge for our method is the parallax error convolution. We need to integrate over parallax for every source at every proposed set of model parameters. In this section we will use slightly different notation where $\varpi = 1/s$ is the true parallax distance which we are marginalising over, and $\varpi_i$ is the measured parallax for source $i$. The integral we need to evaluate is
\begin{align}
    \int_0^\infty \mathrm{d}s\, &\prob(\varpi_i\,|\,s)\,f(l_i, b_i, G_i, s, \boldsymbol{\psi}) \nonumber\\
    &= \int_0^\infty \mathrm{d}\varpi \,\varpi^{-2} \, \mathcal{N}(\varpi; \varpi_i, \sigma_{\varpi,i})\,f(l_i, b_i, G_i, s, \boldsymbol{\psi}) \nonumber\\
    &\equiv \int_0^\infty \mathrm{d}\varpi\, I(\varpi)
\end{align}
where $\mathcal{N}$ is a normal (Gaussian) distribution with standard deviation $\sigma_{\varpi,i}$ which is the parallax error of source $i$. In Section~\ref{sec:model} we will introduce the absolute magnitude model which is broken into sections with an upper absolute magnitude limit (minimum brightness) for the model. We can then write the integral as a sum of definite integrals
\begin{align}
    \int_0^\infty \mathrm{d}\varpi \,I(\varpi) = \sum_j \,\int_{\varpi_j}^{\varpi_{j+1}} \,\mathrm{d}\varpi \,I(\varpi)
\end{align}
where 
\begin{equation}
    \varpi_j = 10^{(M_j + 10 - G_i)/5}
\end{equation}
and $M_j$ are the magnitude boundaries of the sections. For an unconstrained lower absolute magnitude limit, $\varpi_0=0$. 

We numerically evaluate the integral of each section using the following five step recipe.

\begin{enumerate}
\item Transform into logit-parallax space using the substitution
\begin{ceqn}
\begin{equation}
    x' = \log\left(\frac{\varpi - \varpi_j}{\varpi_{j+1}-\varpi}\right).
\end{equation}
\end{ceqn}
This gives
\begin{ceqn}
\begin{equation}
   \int_{\varpi_j}^{\varpi_{j+1}} \,\mathrm{d}\varpi I(\varpi) =
   \int_{-\infty}^{\infty} \mathrm{d}x' \frac{I}{J}
\end{equation}
\end{ceqn}
where the Jacobian is 
\begin{ceqn}
\begin{equation}
    J = \left|\frac{\partial x'}{\partial \varpi}\right| = \frac{\varpi_{j+1}-\varpi_j}{(\varpi-\varpi_j)(\varpi_{j+1}-\varpi)}.
\end{equation}
\end{ceqn}

\item Find the peak of the logit-transformed integrand by solving
\begin{ceqn}
\begin{equation}
    \frac{\partial}{\partial x'}\left(\frac{I}{J}\right) = 0
    \label{eq:integrand_diff}
\end{equation}
\end{ceqn}
using the bisection algorithm with respect to $\varpi$ initialising at the integration boundaries, $\varpi_j, \varpi_{j+1}$. Transform the parallax of the peak into logit space giving us the mode, $x_0'$.

\item Estimate the width of the peak from the curvature around $x'_0$,
\begin{ceqn}
\begin{equation}
    \sigma_{x'} = \left(\frac{\partial^2I/J}{\partial x'^2}\right)^{-\frac{1}{2}}\bigg\rvert_{x'=x_0'} .
\end{equation}
\end{ceqn}
\item Recenter and rescale via
\begin{ceqn}
\begin{equation}
    x = \frac{x' - x_0'}{\sqrt{2}\sigma_{x'}},
\end{equation}
\end{ceqn}
such that the integrand is approximately $I\sim \exp\left(-x^2\right)$ around the peak.

\item Apply Gauss-Hermite quadrature in $x$-space which gives
\begin{ceqn}
\begin{equation}
    \int_{\varpi_j}^{\varpi_{j+1}} \mathrm{d}\varpi \,I = \sum_k w_k \frac{\sqrt{2}\sigma_{x'}\,I(\varpi(x_k))}{J(\varpi(x_k))} \exp\left(x_k^2\right).
\end{equation}
\end{ceqn}
\end{enumerate}
In our application of the method, we use Gauss-Hermite quadrature with 11 sample points. Increasing the number of sampling points has no appreciable effect on our inferred likelihood. 

A major limitation of this method is that it cannot accurately integrate multimodal integrands. The integrand must be unimodal such that we can integrate around the single peak. We will discuss the implications of this in Section~\ref{sec:model} when introducing our model.
However, since this is purely a numerical rather than conceptual challenge, we hope future work can improve on our method to allow for more general models to be evaluated and with greater computational efficiency.

\section{Model}
\label{sec:model}

For this work, we only consider high latitudes, $|b|>80^\circ$. There are several reasons for this:
\begin{itemize}
    \item Dust extinction is negligible at high latitudes. Modelling the 3D distribution of dust throughout the Milky Way is a complicated problem on its own \citep{Marshall2006, Green2014}. We quantify the impact of dust extinction on our results in \PaperII.
    \item The in-plane structure of the Milky Way disc is complex with waves, spiral arms and the bar which add vast numbers of free parameters to any spatial model. 
    \item Parallax integration is computationally expensive and scales linearly with the number of sources. By focusing on a subset of \gaia data, we are left with a computationally tractable problem.
\end{itemize}

The aim of this work is to demonstrate how \gaia parallax information can be used to obtain an unbiased model of the Milky Way's stellar content. The vertical distribution at the Solar neighbourhood is a tractable first step in this direction.

The vertical distribution of sources is assumed to be a mixture of three distinct components: thin disc, thick disc and halo. This canonical model has been used for decades since the addition of the second disc component by \citealt{Gilmore1983}. More recent work has shown that -- rather than a dichotomy into thin and thick discs -- there may be a continuous evolution of disc height with stellar metallicity \citep{Bovy2012nothick, Bovy2016disc}. However, since metallicity is not an observable in our sample, we keep to the canonical distinct thin and thick disc model.

Within each component, we assume the spatial and absolute magnitude distributions are separable such that
\begin{equation}
f(l,b,\varpi,M_G) = \sum_{c=\{\mathrm{Tn},\mathrm{Tk},\mathrm{H}\}} w_c\,\nu_c(l,b,\varpi,\boldsymbol{\psi}_\nu) \, \phi_c(M_G,\boldsymbol{\psi}_\phi).
\end{equation}
This is a significant assumption. The thin disc has undergone star formation over long periods and will have correlations between age and metallicity and the vertical and radial dispersion of orbits \citep[e.g. ][]{Ivezic2008, RecioBlanco2014, Martig2016, Snaith2015}. Likewise, the halo is made of multiple stellar populations from in situ star formation and historical merger events \citep[e.g. ][]{Helmi2018, Belokurov2018, Belokurov2020splash}. Nonetheless, we maintain this assumption here in the interests of keeping a simple and tractable model. 

We have deliberately chosen to assume a separable thin disc -- thick disc -- halo Milky Way as this provides a simple and tractable application of our method which we introduced in Section~\ref{sec:method} whilst still returning physically informative parameters. It will be worthwhile applying our method to \gaia data with more detailed model parameterisations. It is beyond the scope of this work because the exact choice of model will depend on the scientific interests of the researcher so we wish to leave this open.

Note that $w_c$ is a free parameter of the model for each component and gives the total number of stars for that component within the given region of the sky and absolute magnitude range.

\subsection{Spatial distributions}

We consider the thin and thick discs to have exponential profiles vertically $\nu_c\propto\exp\left(-\frac{|z|}{h_c}\right)$ similar to previous work \citep[e.g. ][]{Juric2008, Bovy2012, Bovy2016disc}. Other possibilities include $\mathrm{sech}$ or $\mathrm{sech}^2$ profiles, but there is a moderate preference in the data for an exponential profile \citep{Dobbie2020}.  

Since we are only considering high latitudes, we neglect any radial dependence of the vertical density profile. This makes the numerical integral described in Section~\ref{sec:p_integrate} significantly more tractable. The complexity introduced by adding radial dependence is explained in more detail in Appendix~\ref{app:Rintegrand}. The impact of this simplification on the results is tested and quantified in \citetalias{mwtrace2}.

Transforming into heliocentric coordinates $z=s\sin(b)$ and normalising we get the density distribution 
\begin{equation}
    \nu_c(l,b,s) \mathrm{d}V =  \frac{\tan^2(|b|_\mathrm{min})}{2\pi\,h_c^3}s^2 \exp\left(-\frac{|s\sin b|}{h_c}\right) \,\mathrm{d}l\, \mathrm{d}\sin(b) \,\mathrm{d}s,
\end{equation}
where $|b|_\mathrm{min}=80^\circ$ is the on-sky latitude limit of our sample. In \citetalias{mwtrace2} we consider the northern and southern high latitude samples independently, however, in this work we assume that the Galaxy is symmetric above and below the Galactic mid-plane and that the Sun lies perfectly on the plane at $z=0$ pc. This introduces a $\sim20.8$pc systematic offset into our results \citep{Bennett2019}, whose effect on the posterior distributions is quantified in \PaperII.

For the spatial distribution of the halo, we use a spherically symmetric single power law profile centered on the Galactic centre, $\nu_\mathrm{H}(r)\mathrm{d}V \propto r^{-n_\mathrm{H}}$. Many other works also include a free parameter for the halo axis ratio \citep{Juric2008, Mateu2018}, however, as we are only using a narrow window on the sky, there will be limited information to independently constrain the profile and axis ratio of the halo. Furthermore, previous works have either implicitly or explicitly truncated the halo or included a broken power-law profile. The halo profile used in this work is assumed to extend infinitely and as such a normalisation constraint is placed such that $n_\mathrm{H}>3$. This will be in tension with \citealt{Deason2014} and \citealt{Fukushima2019} who find a steeper halo profile beyond $r\sim 50$ kpc and $160$ kpc respectively. This corresponds to a parallax $\varpi<0.02$ mas which is pushing the precision limit of \gaia parallaxes even for bright sources \citep[see Fig. 7 of ][]{Lindegren2021ast}. Therefore, our model should not be significantly sensitive to this shift. 

As we did for the disc profile, we neglect cylindrical radius dependence for the halo by placing all sources at the same projected distance from the Galactic centre as the Sun, $R_0$, such that
\begin{equation}
    r^2 = s^2 \sin^2(b) + R_{0}^2.
\end{equation}
Again this is only valid at high latitudes. The dependence of the source distribution on Galactic longitude can then be neglected. This may lead to systematic biases which are tested in \PaperII.
The spatial model of the halo is given by
\begin{equation}
    \nu_\mathrm{H}(l,b,s) \mathrm{d}V = \mathcal{N}_{\nu_\mathrm{H}} s^2 \left(s^2\sin^2 b + R_{0}^2\right)^{-\frac{n_\mathrm{H}}{2}} \,\mathrm{d}l\, \mathrm{d}\sin b \,\mathrm{d}s
\end{equation}
where
\begin{equation}
    \mathcal{N}_{\nu_\mathrm{H}} = \frac{1}{2\pi}\frac{8\tan^2(b_\mathrm{min})}{\sqrt{\pi}R_{0}^{3-n}} \frac{\Gamma\left(n/2\right)}{\Gamma\left(n/2-3/2\right)}.
\end{equation}
This spatial distribution adds three parameters to the model: the exponential scale height of the thin and thick discs ($h_\mathrm{Tn}$ and $h_\mathrm{Tk}$) and the power-law index of the halo $n_\mathrm{H}$.

\subsection{Luminosity functions}
\label{sec:magmodel}

\begin{figure}
  \centering
  \includegraphics[width=0.495\textwidth]{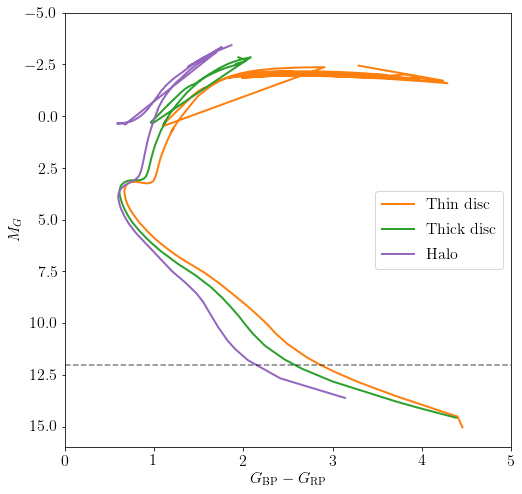}
  \caption[]{HR diagram showing the isochrones used for our mock model of Milky Way sources, with ages $\tau=6.9,7.8,12.5$Gyr and metallicities $\left[\mathrm{Fe/H}\right]=-0.3,-0.7,-1.5$ for the thin disc, thick disc and halo respectively (orange, green and purple). The grey dashed line shows the minimum absolute magnitude of our model -- $M_G=12$.}
   \label{fig:hr_isochrone}
\end{figure}
\begin{figure}
  \centering
  \includegraphics[width=0.495\textwidth]{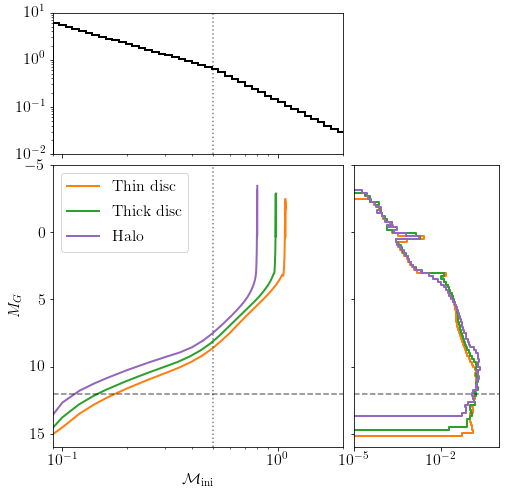}
  \caption[]{The thin disc, thick disc and halo isochrones (orange, green and purple) are used to transform a mock sample of stars from initial mass ($\mathcal{M}_\mathrm{ini}$) to absolute magnitude ($M_G$). The initial mass (top panel) is drawn from a Kroupa IMF \citep{Kroupa2001} with $\mathcal{M}_\mathrm{ini}>0.09\mathrm{M}_\odot$ where the vertical grey dotted line is the break mass $0.5\,\mathrm{M}_\odot$. This produces the absolute magnitude distribution shown in the right hand panel. The horizontal grey-dashed line shows the maximum absolute magnitude as our model only includes sources with $M_G<12$.}
   \label{fig:massmag_isochrone}
\end{figure}
\begin{figure}
  \centering
\includegraphics[width=0.495\textwidth]{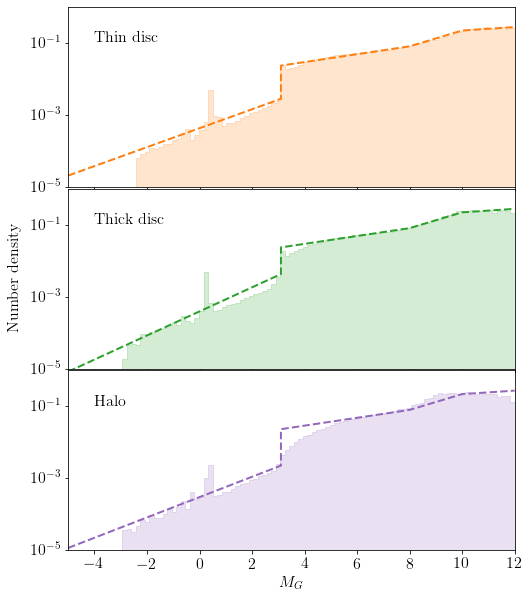}
  \caption[]{The mock distributions produced by transforming the Kroupa IMF through isochrones from Fig.~\ref{fig:hr_isochrone} (shaded histograms) are fit with the approximate absolute magnitude distribution used for the model (dashed lines).}
   \label{fig:magfit_isochrone}
\end{figure}

The luminosity distribution function of stars in the Milky Way is an intricate function of the star formation history, accretion history and dynamical evolution of the Galaxy. The aim of this work is to derive the spatial distribution of sources in the Galaxy - independent of stellar populations - and so the magnitude distribution is only included in order to formally account for the survey selection function. In this section, we will explain how to derive an adequate parameterisation for the luminosity function for each Milky Way component.

Each of the three Milky Way components is assumed to be a single mono-age, mono-abundance stellar population.
Using the results of \citealt{Kilic2017} from white dwarf populations, the ages used for the thin disc, thick disc and halo are $6.9$ Gyr, $7.8$ Gyr and $12.5$ Gyr respectively. Using SDSS spectroscopy, \citealt{Ivezic2008} derived halo and thick disc metallicities of $\mathrm{\left[Fe/H\right]}=-1.5, -0.7$ respectively, whilst \citealt{RecioBlanco2014} used the \gaia-ESO survey~\citep{Gilmore2012} to find the thin disc metallicity fell in the range $[-0.8, 0.2]$ and the thick disc between $[-1.0,-0.25]$. Combining these results, we assume the thin disc, thick disc and halo have metallicities of $-0.3,-0.7$ and $-1.5$. The HR diagram in Fig.~\ref{fig:hr_isochrone} shows the three isochrones which are taken from PARSEC v1.2s \citep{PARSEC2012, Tang2014, Chen2014, Chen2015}.

We then draw a random sample from the broken power law initial mass function (IMF) of \citealt{Kroupa2001} for initial masses greater than $0.09\mathrm{M}_\odot$ with ${\mathcal{M}_\mathrm{ini}\sim \mathcal{M}_\mathrm{ini}^{-1.3}}$ for ${\mathcal{M}_\mathrm{ini}<0.5\,\mathrm{M}_\odot}$ and ${\mathcal{M}_\mathrm{ini}\sim \mathcal{M}_\mathrm{ini}^{-2.3}}$ otherwise. This is shown in the top panel of Fig.~\ref{fig:massmag_isochrone}. The individual component isochrones, shown in the middle panel, are then used to transform the IMF into an absolute magnitude distribution which is shown in the right hand panel. This sample is not used as our mock catalogue, it is only for deriving our model absolute magnitude distribution.

\begin{table*}
\begin{tabular}{c c c c c} 
 \hline
 Component & Parameter & Prior & Transformation & Bounds \\ [0.5ex] 
 \hline\hline
 Thin disc & $w$ & $\mathrm{Dirichlet}(a=2)$ & $\log(w)$ & [-10,50] \\ 
 & $h_\mathrm{Tn}$ & $\mathrm{U}[0.1,0.6]$ & $\mathrm{logit}\left(\frac{h-0.1}{0.6-0.1}\right)$ & [-10,10] \\
 &$f_D$ & $\mathrm{U}[0,1]$ & $\mathrm{logit}(f_D)$ & [-10,10]  \\
 \hline\hline
 Thick disc &$w$ & $\mathrm{Dirichlet}(a=2)$ & $\log(w)$ & [-10,50] \\ 
 &$h_\mathrm{Tk}$ & $\mathrm{U}[0.6,3.0]$ & $\mathrm{logit}\left(\frac{h-0.6}{3.0-0.6}\right)$ & [-10,10] \\
 & $f_D$ & $\mathrm{U}[0,1]$ & $\mathrm{logit}(f_D)$ & [-10,10]   \\
 \hline\hline
 Halo & $w$ & $\mathrm{Dirichlet}(a=2)$ & $\log(w)$ & [-10,50] \\ 
 & $n_\mathrm{H}$ & $\mathrm{U}[3,7.3]$ & $\mathrm{logit}\left(\frac{h-3}{7.3-3}\right)$ & [-10,10] \\
 &$f_D$ & $\mathrm{U}[0,1]$ & $\mathrm{logit}(f_D)$ & [-10,10]   \\
 \hline\hline
 Shared &$\alpha_1$ & $-\alpha_1 \sim \log\mathrm{U}[\mathrm{e}^{-5}, \mathrm{e}^3]$ & $\log(-\alpha_1)$ & [-5,3] \\
 &$\alpha_2$ & $-\alpha_2 \sim \log\mathrm{U}[\mathrm{e}^{-5}, \mathrm{e}^3]$ & $\log(-\alpha_2)$ & [-5,3] \\
 \hline\hline
\end{tabular}
\caption{The 11 free parameters used to model the spatial and absolute magnitude distributions of sources along with their priors. The method fits directly to the parameters under the given transformations where logistic priors are also included to correct for the logit transform. The bounds are applied to the transformed parameters for numerical stability of the optimization.}
\label{tab:priors}
\end{table*}

The absolute magnitude distributions of the three components from the right hand panel of Fig.~\ref{fig:massmag_isochrone} are shown as shaded histograms in Fig.~\ref{fig:magfit_isochrone}. They are made up of four regimes. At the bright end ($M_G\lesssim 3$, sources evolve much faster along the giant branch than the main sequence (MS), generating a sharp drop at the turn-off above which the number density of sources falls quickly aside from a spike at the red clump ($M_G\sim0$). The MS has three components, a relatively shallow upper sequence for $M_G\sim[3, 7]$, a steeper section for $M_G\sim[7,9]$ where the slope of the main sequence in Fig.~\ref{fig:hr_isochrone} shifts which is also around the power-law break of the IMF (we'll refer to this section as the `gap'), and a very flat lower MS for $M_G \gtrsim 9$. Sources continue fainter to the brown dwarf regime; however, stellar models in these regions of parameter space are poorly constrained by observations as there are few stars this dim yet bright enough for current observatories. For this reason, we only consider sources with $M_G>12$ in this work. This will be especially beneficial when we model the \gaia data in \citetalias{mwtrace2} as the majority of sources with spurious astrometric solutions as classified by \citealt{Rybizki2021} and \citealt{Smart2021} have absolute magnitudes fainter than $M_G=12$.

Each component of the absolute magnitude distribution is modelled by an exponential distribution. Here we state the parameterisation, however, a full derivation of the absolute magnitude profile is given in Appendix~\ref{app:mag_dist}. The absolute magnitude is drawn from a broken exponential distribution,
\begin{equation}
    M_G \sim \exp(-\alpha M_G),
\end{equation}
with four components
\begin{align}
    \alpha = \begin{cases}
    \alpha_1 & 9<M_G<12 \hspace{1cm}\mathrm{(Lower\, MS)} \\
    \alpha_g & 7<M_G<9 \hspace{1.15cm}\mathrm{(MS,`gap')} \\
    \alpha_2 & M_\mathrm{TO}<M_G<7 \hspace{0.7cm}\mathrm{(Upper\,MS)} \\
    \alpha_G & M_G<M_\mathrm{TO} \hspace{1.25cm}\mathrm{(Giants)}
    \end{cases}
\end{align}
where $M_\mathrm{TO}$ is the turn-off magnitude. 

The distribution is continuous everywhere except from at the turnoff where the discontinuous change in the gradient of the magnitude-initial mass relation leads to a discontinuity in the magnitude distribution. Continuity conditions at $M_G=7, 9$ constrain the exponential profile $\alpha_g$ and the normalisation $A_g$ of the gap profile.

The full magnitude distribution is given by 
\begin{align}
    f(M)\mathrm{d}M &=\begin{cases}
    (1-f_G)\mathcal{N}_D\frac{1}{a_1}&\\\hspace{1cm}
    \times\exp\left(-\alpha_1 (M-M_\mathrm{MS})\right)\mathrm{d}M 
    & 9<M<12\\
    (1-f_G)\mathcal{N}_DA_g&\\\hspace{1cm}
    \times\exp\left(-\alpha_g (M-M_\mathrm{MS})\right)\mathrm{d}M 
    & 7<M<9\\
    (1-f_G)\mathcal{N}_D\frac{1}{a_2}&\\\hspace{1cm}
    \times\exp\left(-\alpha_2 (M-M_\mathrm{MS})\right)\mathrm{d}M 
    & M_\mathrm{TO}<M<7 \\
    f_G \mathcal{N}_G &\\\hspace{1cm}
    \times\exp(-\alpha_G (M-M_\mathrm{TO}))\mathrm{d}M
    \hspace{-1cm}& M<M_\mathrm{TO}
    \end{cases}
\end{align}  
where $\mathcal{N}_D$ and $\mathcal{N}_G$ are the normalisations of the dwarf and giant magnitude distributions respectively. $M_\mathrm{MS}=8$ is the magnitude of the transition from the lower to upper main sequence.


The magnitude distribution introduces five parameters: $\alpha_1$, $\alpha_2$, $M_\mathrm{TO}$, $\alpha_G$ and $f_G$, the fraction of the population which are giants, which constrains the size of the discontinuity at the turn-off. We could fix all parameters using the IMF-isochrone sample just constructed. However, this is only an approximate representation of the magnitude distribution which may introduce large systematics. To avoid this problem, we free up $\alpha_1$, $\alpha_2$ and $f_G$ to be constrained by the real data. $\alpha_1$ and $\alpha_2$ are assumed to be the same for all populations as the MS is dominated by older stars which show a similar distribution independent of population parameters.

The position of the turn-off, $M_\mathrm{TO}$, defines a discontinuity for the model. Depending on the location of individual sources in relation to the turnoff, this can generate sample-dependent local optima in the likelihood space which is challenging for optimization. For this reason, we fix $M_\mathrm{TO}=3.1$ for all models and address the implications of this in \PaperII. $\alpha_G$ has a strong degeneracy with $f_G$ as both control the number of sources at bright magnitudes. We also fix $\alpha_G$ in all optimizations to avoid this degeneracy to values which are discussed in Section~\ref{sec:mock}. All free parameters are listed in Table~\ref{tab:priors} with their respective components.

This fully defines the model which we fit to the \gaia data. In total, there are 11 free parameters of the model.

\begin{figure}
  \centering
\includegraphics[width=0.495\textwidth]{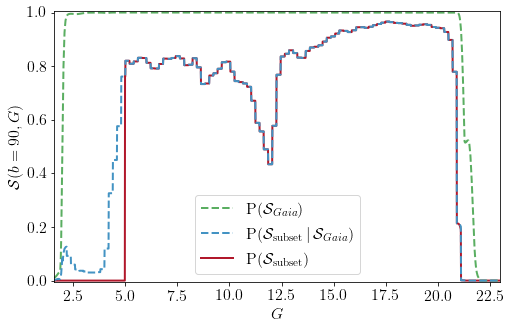}
  \caption[]{The selection function probability at $b=90^\circ$ for the \gaia EDR3 source catalogue (green dashed) drops off at bright magnitudes ($G<2$, due to CCD over-saturation) and faint magnitudes ($G\gtrsim 21$) however remains high across the rest of apparent magnitude space. The \gaia EDR3 astrometry with $\mathrm{RUWE}<1.4$ relative selection function (blue dashed) is more restrictive over the entire magnitude range and dominates the total selection function (red solid). The cut-off at $G<5$ is deliberately imposed to remove regions of apparent magnitude with poor astrometry calibration.}
   \label{fig:sf_g}
\end{figure}

\renewcommand{\arraystretch}{1.5}
\begin{table*}
\begin{tabular}{c c c c c c} 
 \hline
 Component & Parameter & Input & Full & SF & SF \& $\sigma_\varpi$ \\ [0.5ex] \hline\hline
Thin disc& $\log_{10}(w)$ & ${4.0792}$ & ${4.0700}
        _{-0.0191}
        ^{+0.0194}$ & ${4.0637}
        _{-0.0359}
        ^{+0.0331}$ & ${3.9816}
        _{-0.0657}
        ^{+0.0586}$\\ 
& $h_\mathrm{Tn}$ & ${0.300}$ & ${0.301}
        _{-0.006}
        ^{+0.007}$ & ${0.301}
        _{-0.010}
        ^{+0.010}$ & ${0.281}
        _{-0.015}
        ^{+0.015}$\\ 
& $f_G$ & ${4.50}\times 10^{-3}$ & ${3.73}
        _{-0.99}
        ^{+1.00}\times 10^{-3}$ & ${3.91}
        _{-1.22}
        ^{+1.23}\times 10^{-3}$ & ${3.76}
        _{-1.30}
        ^{+1.43}\times 10^{-3}$\\ 
& $M_\mathrm{TO}$ & 3.1 & & & \\& $\alpha_3$ & -0.6 & & & \\\hline\hline 
Thick disc& $\log_{10}(w)$ & ${4.6335}$ & ${4.6249}
        _{-0.0050}
        ^{+0.0051}$ & ${4.6253}
        _{-0.0093}
        ^{+0.0092}$ & ${4.6221}
        _{-0.0198}
        ^{+0.0200}$\\ 
& $h_\mathrm{Tk}$ & ${0.900}$ & ${0.891}
        _{-0.011}
        ^{+0.012}$ & ${0.884}
        _{-0.028}
        ^{+0.029}$ & ${0.812}
        _{-0.045}
        ^{+0.052}$\\ 
& $f_G$ & ${5.40}\times 10^{-3}$ & ${5.76}
        _{-0.52}
        ^{+0.54}\times 10^{-3}$ & ${5.80}
        _{-0.60}
        ^{+0.64}\times 10^{-3}$ & ${5.83}
        _{-0.66}
        ^{+0.69}\times 10^{-3}$\\ 
& $M_\mathrm{TO}$ & 3.1 & & & \\& $\alpha_3$ & -0.77 & & & \\\hline\hline 
Halo& $\log_{10}(w)$ & ${5.9754}$ & ${5.9759}
        _{-0.0005}
        ^{+0.0005}$ & ${5.9662}
        _{-0.0105}
        ^{+0.0106}$ & ${5.9450}
        _{-0.0229}
        ^{+0.0247}$\\ 
& $n_\mathrm{H}$ & ${3.740}$ & ${3.745}
        _{-0.001}
        ^{+0.001}$ & ${3.753}
        _{-0.020}
        ^{+0.020}$ & ${3.812}
        _{-0.066}
        ^{+0.068}$\\ 
& $f_G$ & ${3.50}\times 10^{-3}$ & ${3.47}
        _{-0.06}
        ^{+0.06}\times 10^{-3}$ & ${3.49}
        _{-0.09}
        ^{+0.10}\times 10^{-3}$ & ${3.48}
        _{-0.15}
        ^{+0.15}\times 10^{-3}$\\ 
& $M_\mathrm{TO}$ & 3.1 & & & \\& $\alpha_3$ & -0.64 & & & \\\hline\hline 
Shared& $\alpha_1$ & ${-0.1100}$ & ${-0.1109}
        _{-0.0004}
        ^{+0.0003}$ & ${-0.1094}
        _{-0.0015}
        ^{+0.0014}$ & ${-0.1098}
        _{-0.0020}
        ^{+0.0020}$\\ 
& $\alpha_2$ & ${-0.2500}$ & ${-0.2524}
        _{-0.0019}
        ^{+0.0020}$ & ${-0.2534}
        _{-0.0046}
        ^{+0.0045}$ & ${-0.2521}
        _{-0.0084}
        ^{+0.0089}$\\ 
\hline\hline 
\end{tabular}
\caption{The input parameters for the mock sample catalogue generation and the results of the fit to the data are shown when using the full sample with no observational errors ("Full"), the selection function with no observational errors ("SF") and the sample with both the selection function and the added parallax errors ("SF \& $\sigma_\varpi"$). For all parameters we provide the median and $16^\mathrm{th}$ and $84^\mathrm{th}$ percentile uncertainties. 
}
\label{tab:mockparams}
\end{table*}

\subsection{Selection Function}
\label{sec:sf}

A major obstacle to using a catalogue of sources to fit a distribution is the selection function. Many surveys have complex and unknown observation limitations which are a strong function of observatory properties and observing conditions. \gaia is no exception due in part to the complexity of the scanning law \citep{CoGI, CoGIII}. 

In most previous works, the sample is either assumed to be magnitude complete to some limit \citep[e.g.][]{Juric2008, Bilir2006sdss, Ak2008}, or the sample is bright and nearby for which there are larger, complete catalogues against which the selection function has been estimated \citep[e.g.][]{Bovy2017, Mateu2018, Bennett2019}. \gaia is neither complete in position on the sky or apparent magnitude, nor is there a larger, more complete sample against which to compare the \gaia catalogue.

Fortunately, a solution for the \gaia source catalogue selection function has been developed and applied to \gaia DR2 \citep{CoGII}. Appendix~A of \citealt{CoGV} provides a simple extension to model the selection function of the \gaia EDR3 source catalogue using the nominal EDR3 scanning law. This may have some limitations in crowded regions due to changes in \gaia's data processing pipeline. However, since we are only considering high latitude fields, it should be sufficient for our purposes. The selection probability as a function of apparent magnitude for ${b=90^\circ}$ is given by the green dashed line in Fig.~\ref{fig:sf_g} showing that the source catalogue is nearly complete for $3<G<21$.



Given the source catalogue selection function, the selection functions of subsets can be estimated by comparison \citep{Boubert2021, CoGV}. In \PaperII, we will use the \gaia astrometry catalogue with $\mathrm{RUWE}<1.4$ where apparent $G$-band magnitude is available. The selection function for this dataset is given by the product of the source catalogue and subset selection functions
\begin{equation}
    \mathcal{S}_\mathrm{subset}(l, b, G) =  \prob(\mathcal{S}_\mathrm{subset} \,|\, \mathcal{S}_\mathrm{source}, l, b, G) \, \prob(\mathcal{S}_\mathrm{source} \,|\, l, b, G)   
    \label{eq:sf_definition}
\end{equation}
where $\prob(\mathcal{S}_\mathrm{source} \,|\, l, b, G)$ is the probability of selection in the \gaia source catalogue with published $G$ and ${\prob(\mathcal{S}_\mathrm{subset} \,|\, \mathcal{S}_\gaia, l, b, G)}$ is the probability of an object in the source catalogue having published parallax with $\mathrm{RUWE}<1.4$, modelled in \citealt{CoGV}, as a function of $G$ and position on the sky only. When fitting the model parameters to data, Eq.~\ref{eq:sf_definition} is substituted into Eq.~\ref{eq:likelihood}.

The results are applied in 0.2mag bins in $G$ in $\textsc{nside}=64$ HEALPix pixels \citep{Gorski2005} across the sky. The selection probability for $b=90^\circ$ is given by the red line in Fig.~\ref{fig:sf_g}.
Due to the challenges of modelling sources which saturate the \gaia CCDs at the bright end of the magnitude distribution we use a selection function which truncates at $G=5$. Our sample will also only include those sources with $G>5$.

\section{Mock}
\label{sec:mock}

To test and demonstrate the efficacy of the method, we generate a mock catalogue from our model with realistic parameters. Information on the true parameters is then removed, the \gaia selection function and \gaia-like parallax uncertainties are applied, and we attempt to infer the input parameters from the mock sample. We note that this only tests the method. Because the data is drawn from the same model which is being refit, any inconsistencies between the model and true Milky Way distribution of stars do not show up here. These inconsistencies are discussed, tested and quantified in \PaperII.

\subsection{Input parameters}

Parameters for the scale heights and power law indices of the discs and halo respectively are taken from the literature. For the thin disc $h_\mathrm{Tn}=300$pc and thick disc $h_\mathrm{Tk}=900$pc \citep{Juric2008}. The power law index used is $n_\mathrm{H}=3.74$ from \citealt{Fukushima2019}.

The relative stellar mass density of the discs is $\rho_\mathrm{Tk}/\rho_\mathrm{Tn}=0.12$ and $\rho_\mathrm{H}/\rho_\mathrm{Tn}=0.005$ \citep{Juric2008}. Instead of local mass density, our model fits the total number of sources in each component with $|b|>80^\circ$. To convert mass density to number density in the Solar neighbourhood, we divide by the mean mass of a star. The mean mass is estimated using the IMF-isochrone sample in Section~\ref{sec:magmodel} as $\mathcal{M} \sim 0.413,0.369,0.308\,\mathrm{M}_\odot$ for the thin disc, thick disc and halo respectively. We then divide the number density by the value of the normalised component at $s=0$ to get the total number of sources in each component. The result is that $w_\mathrm{Tn}/w_\mathrm{Tk}=0.275$ and $w_\mathrm{Tn}/w_\mathrm{H}=0.0127$. The halo dominates the total counts, because our observing volume is a cone with $|b|>80^\circ$. This significantly reduces the relative contribution from the disc to the sample.

The absolute magnitude distributions for each Milky Way component are shown by the shaded histograms in Fig~\ref{fig:magfit_isochrone}. To estimate magnitude parameters for the luminosity function described in Section~\ref{sec:magmodel}, we directly fit the parameters to the magnitude distributions. For each component, the turn-off magnitude is at $M_G \sim 3.1$. $f_G$ is approximated from the ratio of sources with $G<3.1$ to those with $G>3.1$. For $G<3.1$ we fit a power law profile to each component independently using the Poisson likelihood function from Eq.~\ref{eq:poissonlike}. This gives $\alpha_3 = -0.60, -0.77, -0.64$ and $f_G=0.0045, 0.0054, 0.0035$  for the thin disc, thick disc and halo respectively. 

The lower main sequence is dominated by old, long-lived stars which evolve slowly on the HR diagram. Therefore, we assume that the main sequence profiles are similar between different Milky Way components such that the values of $\alpha_1, \alpha_2$ are shared between profiles. We draw a sample of sources from each of the components according to the component's respective weight and fit the main sequence profiles to the sources with $G>3.1$, which gives $\alpha_1=-0.12$, $\alpha_2=-0.26$. The dashed lines in Fig.~\ref{fig:magfit_isochrone} give the absolute magnitude distributions implied by the parameters we have just derived. 

All selected and evaluated parameter values are listed as `Input' in Table~\ref{tab:mockparams}.

\begin{figure*}
  \centering
  \includegraphics[width=\textwidth]{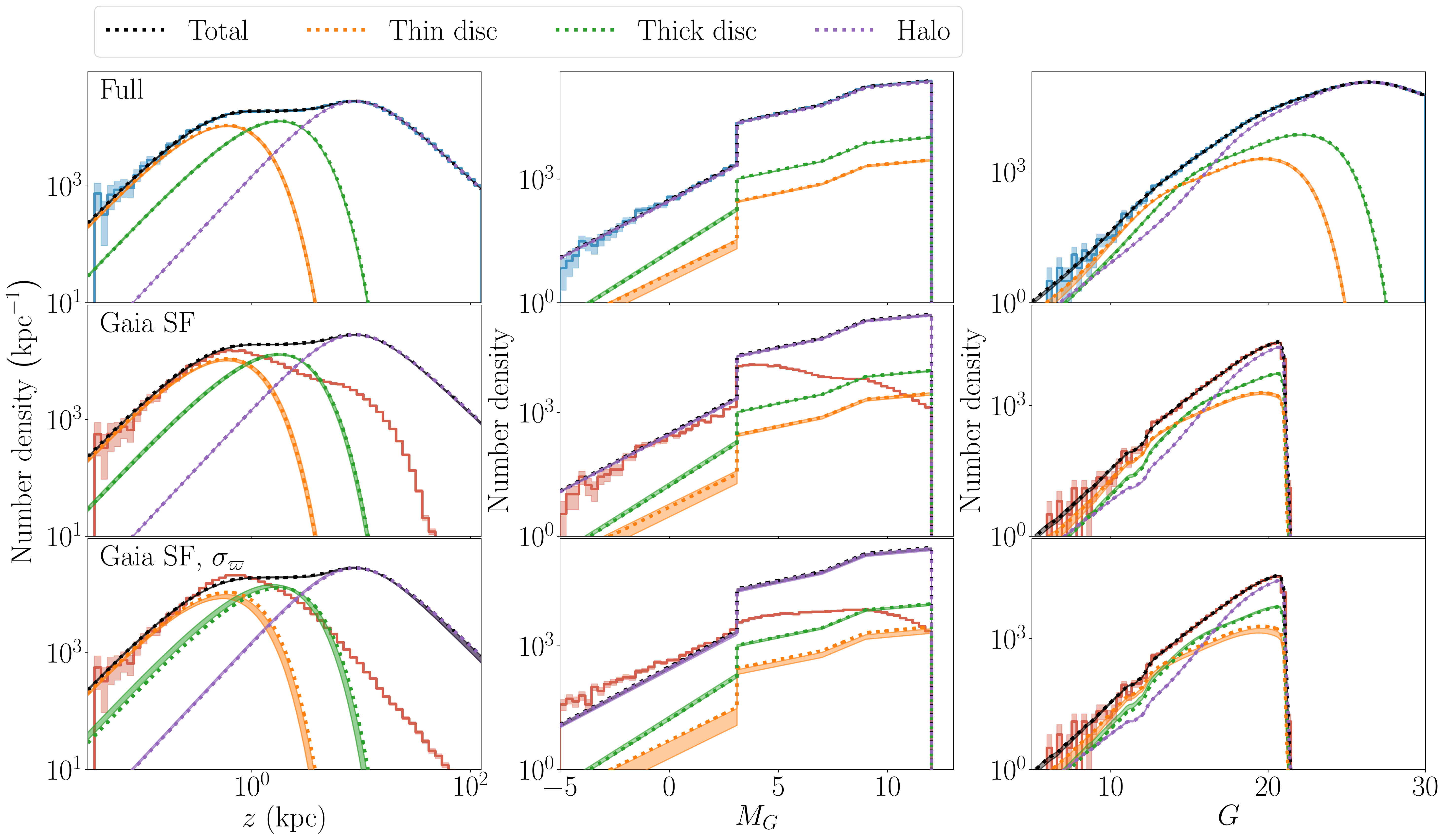}
  \caption[]{The posterior distribution of fits to the mock sample are shown by the shaded regions for the thin disc (orange), thick disc (green), halo (purple) and the sum total (black) as a function of vertical height ($z$, left), absolute magnitude ($M_G$, middle) and apparent magnitude ($G$, right). Blue histograms in the top row show the full sample which the model fit perfectly cut through. The red histograms in the middle and bottom rows show the distribution of SF selected samples with the bottom row showing the distribution of $z=\sin(b)/\varpi$ and $M_G=G+5\log_{10}(\varpi)-10$, demonstrating the impact of parallax uncertainties on measured quantities. The posteriors agree extremely well with the ground truth shown by the dotted lines in all panels. This is true when fitting to the full sample (top), the SF-limited sample (middle) and the SF-limited sample with measured parallaxes sampled from their error distributions (bottom). The posterior distributions are evaluated by randomly selecting $100$ samples from the MCMC posteriors and taking the $16^\mathrm{th}-84^\mathrm{th}$ percentile range. In several cases, particularly for the ``Full'' fits in the top row, the posterior is so tight that the distribution appears as a line in the figure.}
   \label{fig:mock_zM}
\end{figure*}

\subsection{Parallax error}
\label{sec:asf}

To generate a realistic mock, we also need to sample measurement uncertainties. Since the \gaia astrometry was fit using an iterative linear regression process, the covariance may be estimated from information theory (neglecting excess noise) using only the scanning law and individual observation centroid uncertainties. This process is performed in \citealt{CoGIV} for \gaia DR2 and we use the \gaia EDR3 nominal scanning law to extend this to the EDR3 baseline.

The covariance estimates break down for sources with significant excess noise, such as in heavily crowded regions and for sources with intrinsic astrometric variability like binaries. Since we will only consider sources with $|b|>80^\circ$, crowding is negligible. By focusing on the sample with $\mathrm{RUWE}<1.4$, we expect to have removed sources with observable binary motion.

\begin{figure*}
  \centering
  \includegraphics[width=\textwidth]{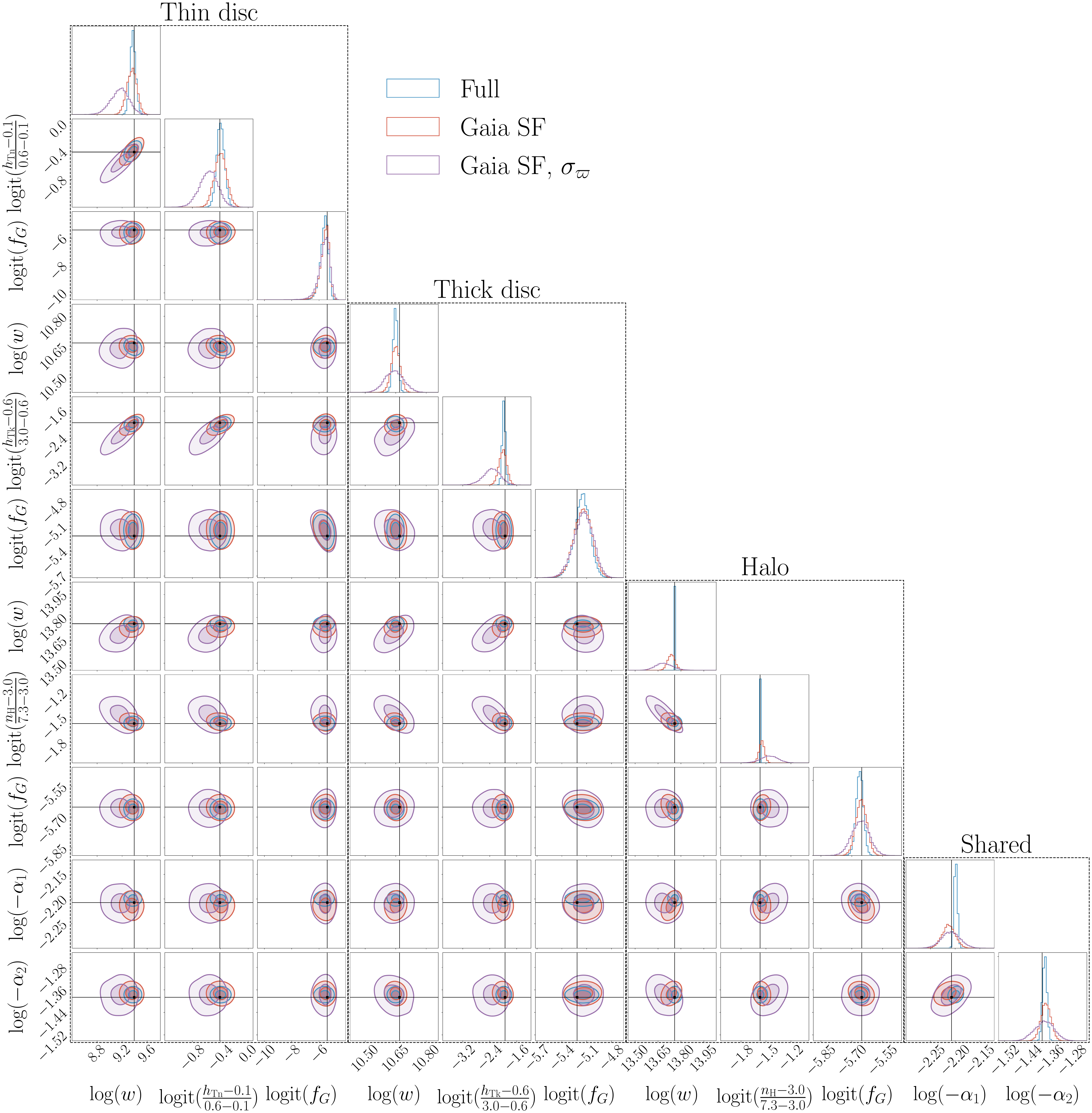}
  \caption[]{The posterior distributions for all mock samples are shown as a function of transformed parameters which are fit to the data. The Full sample fits (blue), SF sample (red) and SF with parallax error (purple) all show strong agreement with one another and the input parameters (black lines). The enhancement of the statistical uncertainty by introducing parallax error can clearly be seen by the increased spread of the posterior for the purple contours.}
   \label{fig:mock_corner_f}
\end{figure*}

\subsection{Mock samples}

A sample of one million sources with distance, latitude and absolute magnitude is drawn from the model using MCMC sampling \citep{ForemanMackey2013}. Since all sources are assumed to be at the projected distance from the Galactic centre of the Sun, the full model is Galactic longitude-independent so the longitude is drawn from a uniform distribution $l\sim\mathrm{U}[0,2\pi]$.
The distribution of drawn sources as a function of distance from the Galactic disc and absolute magnitude is given by the blue histograms in the top panels of Fig.~\ref{fig:mock_zM}. 

The selection function probability is evaluated for all sources given their position on the sky and apparent magnitude as described in Section~\ref{sec:sf}. To generate the mock \gaia astrometry with $\mathrm{RUWE}<1.4$ sample, the event of a source being included is drawn from a Bernoulli distribution with the given selection probability ${\mathrm{S}_i \sim \mathrm{Bernoulli}(\mathcal{S}(l_i, b_i, G_i))}$ where ${\mathrm{S}_i = 0,1}$. Of the $1,000,000$ source in the full sample, 73,132 survive the selection cuts, shown by the red histograms in the middle and bottom panels of Fig.~\ref{fig:mock_zM}.

Parallax error is evaluated from the Astrometric Spread Function described in Section~\ref{sec:asf}. The observed parallax is drawn from a Gaussian distribution with the given error for each source $\varpi\sim \mathcal{N}(1/s, \sigma_\varpi)$. The red histograms in the bottom panels of Fig.~\ref{fig:mock_zM} show the distribution of measured $z=\sin(b)/\varpi$, ${M_G=G-10+5\log_{10}(\varpi/\mathrm{mas})}$ after sampling $\varpi$ from the parallax error. This significantly affects the distributions, demonstrating the importance of properly accounting for parallax uncertainty when modelling the structure of the Milky Way from \gaia data.

This produces three samples which can each be used to independently fit the model parameters demonstrating each stage of the method:
\begin{enumerate}
    \item Full sample fit with Eq.~\ref{eq:poissonlike}: $l^i, b^i, si, G^i \,\forall\, i$,
    \item SF sample fit with Eq.~\ref{eq:sflike}: $l^i, b^i, s^i, G^i \,\forall\,i$ where $S_i=1$,
    \item SF \& $\sigma_\varpi$ fit with Eq.~\ref{eq:likelihood}: $l^i, b^i, \varpi^i, G^i \,\forall\,i$ where $S_i=1$. 
\end{enumerate}
To be clear, in sample (iii) the selection function is not dependent on measured parallax or parallax error as discussed in Section~\ref{sec:method}. We simply mean that the selection function is applied and parallax error on sources is also included. Samples (ii) and (iii) contain the exact same subset of sources from the mock catalogue. Sample (ii) has no parallax error, whilst measured parallaxes in (iii) have been drawn from the parallax uncertainties.

\section{Parameter Inference}
\label{sec:infer}

In this section we will use the method introduced in Section~\ref{sec:method} to fit the model parameters to the three mock samples described in Section~\ref{sec:mock}.

\subsection{Priors}

Priors for all free parameters of the fits are given in Table~\ref{tab:priors}. As is common with mixture model fits to density distributions, the likelihood space is strongly multi-modal. For the thin and thick discs there is of course a complete degeneracy where the components can be switched, but there are also problematic modes where, for example, a single component is expanded to fit the full data-set, whilst remaining components are suppressed.

Priors are chosen specifically to avoid local optima in the model. All weights are assumed to be drawn from a Dirichlet distribution with $a=2$ to remove modes where any component is completely suppressed relative to the others. To avoid the disc degeneracy, the possible disc scale heights are limited to non-overlapping ranges with $h_\mathrm{Tn}\sim\mathrm{U}[0.1\mathrm{kpc},0.6\mathrm{kpc}]$ and $h_\mathrm{Tk}\sim\mathrm{U}[0.6\mathrm{kpc},3.0\mathrm{kpc}]$. The power-law index of the halo is also limited to $n_\mathrm{H}\sim\mathrm{U}[3.0, 7.3]$ as $n_H<3.0$ would produce an unnormalised halo and $n_H>7.3$ produces an incredibly steep halo profile which can mimic the exponential discs (for $n_H=7.3$ the mean halo source distance is the same as an exponential profile with $h=3.0$ kpc). 

For numerical stability, the fits are made on the transformed parameters where transformations are given in Table~\ref{tab:priors}. The transformations scale parameters to the range $[-\infty, \infty]$ in all cases. For logit transformed parameters, we include a logistic prior in logit space which is equivalent to a uniform prior in untransformed space. Therefore the logit transformation has no effect on the prior.

The L-BFGS-B algorithm requires boundaries on all parameters which are given in the final column of Table~\ref{tab:priors}. The boundaries are chosen to avoid regions of parameter space which suffer from numerical precision issues. None of the parameter posterior distributions push up against the boundaries.

\subsection{Optimization}
\label{sec:optimization}

The likelihood optimization is performed in three stages. All MCMC processes used \textsc{emcee} \citep{ForemanMackey2013}. First, a set of samples is drawn from the parameter priors using MCMC with 44 walkers (this is $4\times$ the number of free parameters in our model), with 100 step burn-in and 100 steps of sampling.
Secondly, ten samples are randomly selected from the prior samples as initialisation for gradient descent using L-BFGS-B \citep{lbfgsb} as implemented in \textsc{scipy}.
Finally, the maximum likelihood estimate with the highest likelihood is taken as the best fit solution. A secondary MCMC process is initialised with 44 walkers drawn from a Gaussian ball around the maximum likelihood estimate with variance of $10^{-10}$ times the boundary width. These walkers were run with the likelihood $\times$ prior for 5000 steps. The latter 2500 steps are used at 5 step intervals as the posterior samples.
This process is used for fitting all mock samples and the real \gaia data in \citetalias{mwtrace2}. 

\subsection{Results}


The `Full' sample posteriors, given by the blue contours in Fig.~\ref{fig:mock_corner_f}, provide tight solutions around the input parameter values which are shown by the black dot. A more quantitative comparison can be made from Table~\ref{tab:mockparams} which shows that the majority of input parameters fall within the $16-84^\mathrm{th}$ percentile range of the posterior distribution. The top panels of Fig.~\ref{fig:mock_zM} compare the ground truth input model, shown with dotted lines, to the refit model, shown by the narrow shaded regions. To produce the shaded posteriors in Fig.~\ref{fig:mock_zM} we draw 100 samples from the posterior parameter distributions and plot the $16^\mathrm{th}-84^\mathrm{th}$ percentile range as a function of $z$, $M_G$ and $G$. The posteriors are so tight in most cases that the shaded regions appear as lines perfectly tracking the input model and the total of the components in black sits exactly on top of the blue histograms which show the distribution of the data in the sample.

The `SF' sample, fit to only 73,132 of the initial one million mock sources, has a significantly less tight constraint around the true parameters, shown by the red contours in Fig~\ref{fig:mock_corner_f}, but the parameters show no significant bias. The fits to the halo parameters are slightly shifted from the true values but all parameters are well within $2\sigma$ of the input so this can be well explained by correlated noise, particularly considering the negative correlation between the halo weight and power-law index, $n_\mathrm{H}$. The red histograms in the middle panels of Fig.~\ref{fig:mock_zM} show the selection-limited sample which drops significantly at large vertical heights and faint apparent magnitudes demonstrating how much the model has to extrapolate using the selection function. Again, the model posteriors sit perfectly on the input model shown by the dotted lines. 

For the apparent magnitude distribution in the middle right panel of Fig.~\ref{fig:mock_zM} we show the model multiplied by the selection function probability. The total model (black) sits perfectly on top of the red sample histograms demonstrating how successfully the model is fit to the data. This distribution will be especially important when analysing fits to the real \gaia data when we cannot directly infer the distance of stars from the Galactic plane or their absolute magnitudes due to significant parallax uncertainties.

The `SF \& $\sigma_\varpi$' posterior, given by the purple contours in Fig.\ref{fig:mock_corner_f}, has significantly enhanced uncertainty compared with the solely SF limited data. This demonstrates how much information is held in the parallax and how information is lost when realistic \gaia parallax uncertainties are included. In spite of this, the input parameters are still recovered with reasonable precision and good accuracy. In the bottom panels of Fig.~\ref{fig:mock_zM}, we can see the posterior samples produce a clearer spread around the input distribution. This time the thin and thick discs have not been perfectly fit within the posteriors however the difference is still small enough to be well explained by statistical noise.

These results have demonstrated that the Poisson-likelihood method accounting for the \gaia selection function and parallax error is a powerful tool for recovering the spatial distribution of sources in the Milky Way. However this only tests the self-consistency of the method; the results may still be susceptible to systematic uncertainties if the model does not represent the real Milky Way.


\section{GUMS}
\label{sec:gums}

So far we have only tested the method on data drawn from the fitted model. But what happens when we attempt to fit our model to a more general and realistic catalogue? To test this we use the \gaia Universe Model Snapshot \citep[GUMS, ][]{Robin2012}, a synthetic Milky Way based on the Besan\c{c}on Galaxy Model \citep{Robin2003} which was developed to test the \gaia data processing pipeline with a realistic population of sources.

The latest GUMS sample is provided with \gaia EDR3 and described in the \gaia documentation\footnote{\url{https://gea.esac.esa.int/archive/documentation/GEDR3/Data_processing/chap_simulated/sec_cu2UM/ssec_cu2starsgal.html}}. We will provide a very brief overview of the key components.

\begin{figure*}
  \centering
  \includegraphics[width=\textwidth]{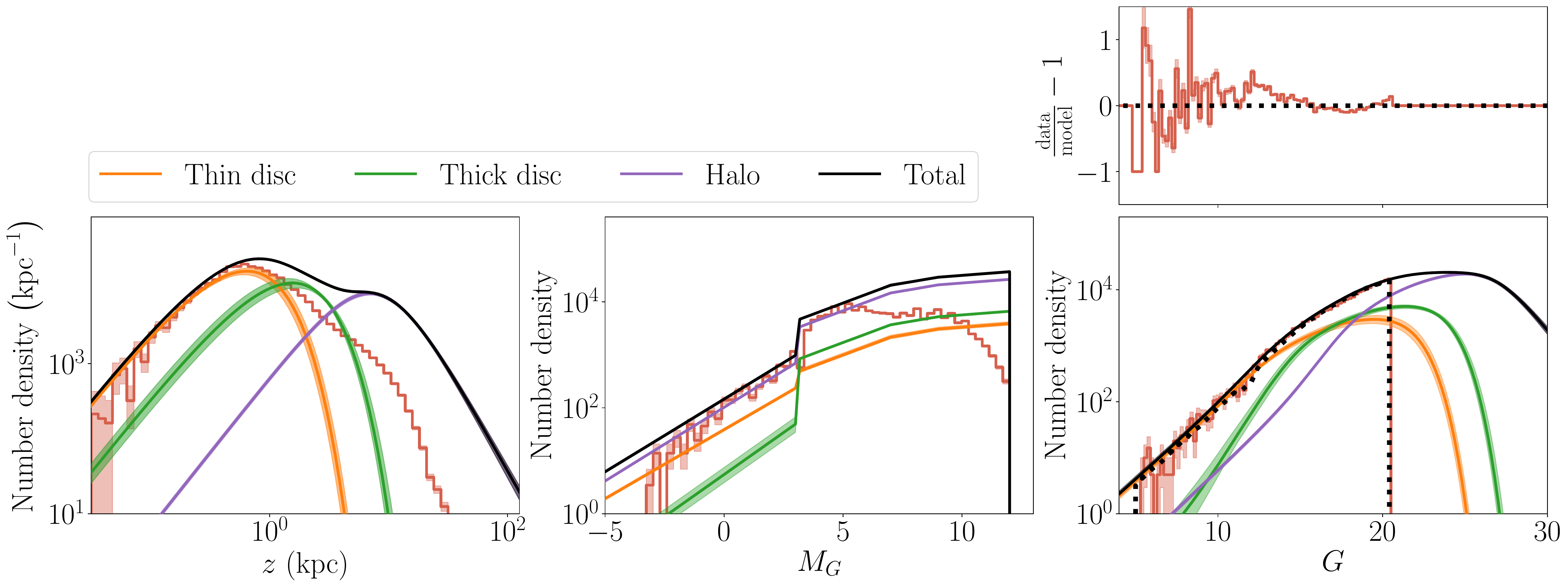}
  \caption[]{\textbf{Bottom row:} Number density of sources in sample (red histograms) and predicted by the model in the thin disc (orange), thick disc (green), halo (purple) and sum total (black). Model shaded regions show the $1^\mathrm{st}$-$99^\mathrm{th}$ percentiles of posterior parameter fits to the data. \textbf{Top right:} Relative residual of the data from the median model fit showing that the model produces a small but significant underestimate of the data at bright magnitudes and overestimate at fainter magnitudes. The shaded red regions show the one standard deviation Poisson uncertainties of the bin counts.}
   \label{fig:gums_zM}
\end{figure*}

The thin disc is composed of seven mono-age populations each contributing a sum of square-exponential radial and vertical profiles to the Milky Way disc. The scale heights of the profiles increase with age, with ages ranging from $0-10$ Gyr. The thick disc is based on the results of \citet{Robin2014} and consists of a sum of two components which are exponentially distributed in Galactocentric radius and reciprocal-$\cosh$-square distributed in vertical height above the midplane with ages $10, 12$ Gyr and scale heights $400, 795$ pc respectively. Flaring is also applied to the disc profiles however this only takes effect for $R>10$ kpc so should not affect our analysis. The spheroidal halo is power-law distributed with $n_\mathrm{H}\sim 3.77$ for Galactocentric distances of $r>>2.2$ kpc and is slightly oblate with $q=0.77$.

The absolute magnitudes of the populations are determined by sampling from an IMF and star formation history and using stellar evolution tracks. The thin discs use a constant star formation history whilst instantaneous bursts of star formation at $10$ and $12$ Gyr are used for the thick disc and $14$ Gyr for the halo.

The sample provided through the \gaia archive includes binary and higher order systems for a significant fraction of stars. For the purposes of this study, we treat all systems as unresolvable with \gaia, i.e. we only include them as a single point source with flux given by the sum of all stars in the system.

We select sources with $b>80^\circ$ in the north and $b<-80^\circ$ in the south from the GUMS catalogue. Due to computational limitations, we work only with a randomly-drawn $10\%$ subsample when testing our method on GUMS. The GUMS sample was cut internally to only include sources with $G<21$ where $G$ was estimated from simple colour relations. The published apparent $G$-band magnitude is computed from the GUMS synthetic spectra such that the originally sharp cut becomes a smooth drop off at $G\sim21$ (Robin, private communication). To avoid this, we cut the sample at $G=20.5$ and set the selection function to $\mathcal{S}=0$ for $G>20.5$. We then produce a \gaia-like mock catalogue by resampling the data from the selection function introduced in Section~\ref{sec:sf}. This produces 47\,027 sources in the north and 49\,508 in the south. The distributions of the north sample as a function of height above the mid-plane, absolute magnitude and apparent magnitude are shown by the red histograms in the lower panels of Fig.~\ref{fig:gums_zM}.

As we did for the mock in Section~\ref{sec:asf}, we also resample a realistic observed parallax measurement for each source from the Astrometric Spread Function \citep{CoGIV}. We then run two fits for each of the north and south samples, one fit to the sample with no parallax error applied and the other with parallax error applied. These are equivalent to fits (ii) and (iii) in Section~\ref{sec:mock}.

The results of the fits to the north sample with parallax error are shown in Fig.~\ref{fig:gums_zM}. The model is only slightly above the data at small $z$ and bright absolute magnitudes which is unsurprising as few sources will have been removed by the selection function in these regions of parameter space. For the apparent magnitude distribution, we also apply the selection function to the total model which produces the black dotted line. The top right panel of the figure shows the relative residuals of the data from the model. At bright magnitudes the residuals are very large but decline to $\sim$a few per cent  for $G \gtrsim 15$. This demonstrates that our model is not flexible enough to accurately reproduce the data, however, at the fainter magnitudes this inaccuracy is small relative to the scale of the model.

\begin{figure*}
  \centering
  \includegraphics[width=\textwidth]{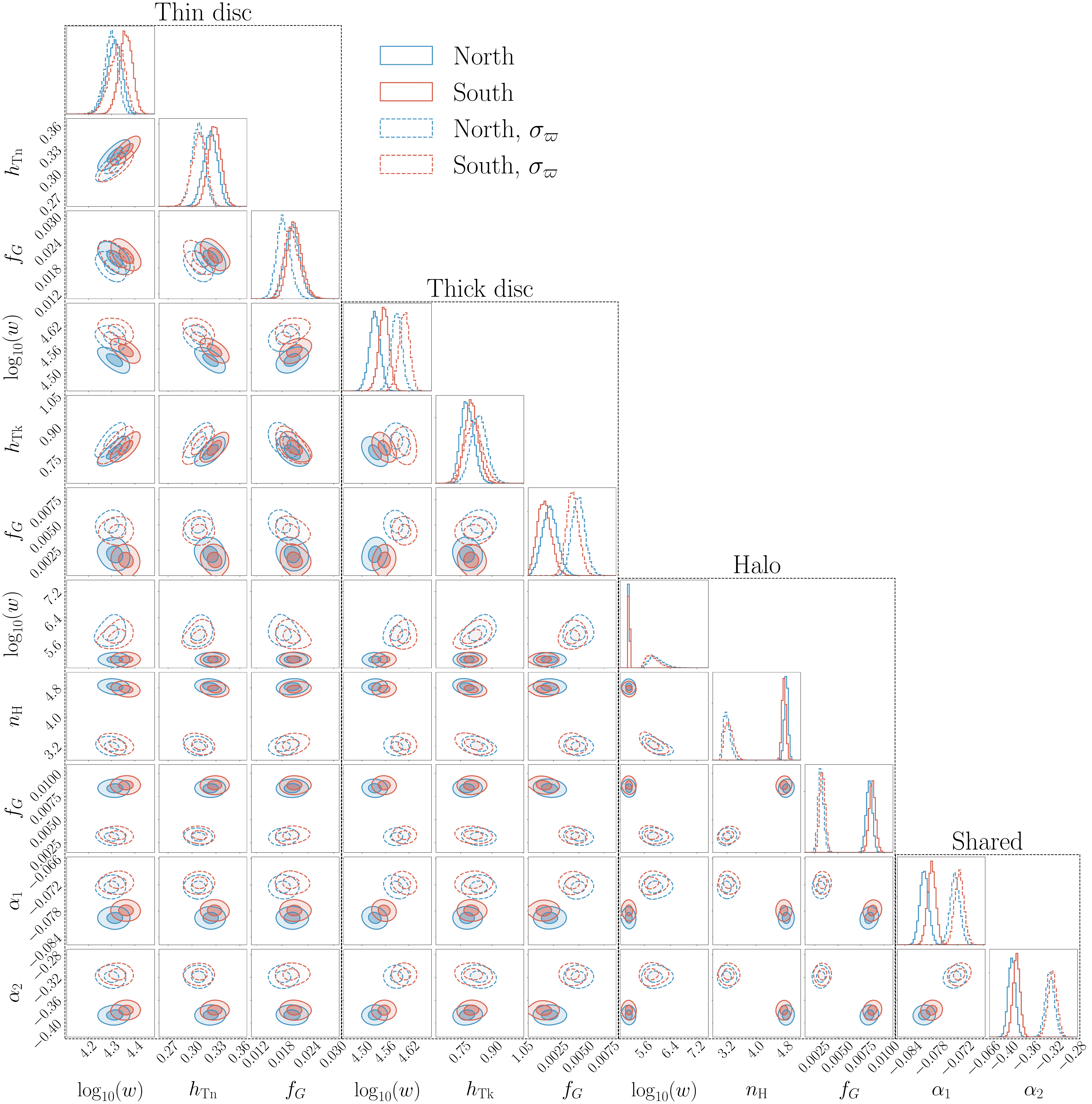}
  \caption[]{Posterior distribution of fits to GUMS data for the north (blue) and south (red) samples show reasonable agreement across all parameters. The fits with precise distances (solid) and parallaxes drawn from uncertainty distributions (dashed) are consistent for disc parameters but significantly disagree for the halo, likely related to a overly simplistic magnitude model.}
   \label{fig:gums_cornerf}
\end{figure*}

The parameter posteriors for all four fits are shown in Fig.~\ref{fig:gums_cornerf}. The first thing to note is that the north and south posteriors, shown by blue and red contours respectively, are consistent with one another for all parameters whether parallax error is applied or not. This demonstrates that we are correctly finding no asymmetry north or south of the Milky Way disc. The glaring problem with our results is that the halo fits are significantly different when exact distances are used (solid contours) and when parallaxes are drawn from uncertainties (dashed contours). The halo distribution peaks at $z\sim 10$ kpc corresponding to a parallax of $\sim 0.1$ mas which is smaller than the parallax uncertainty for sources fainter than $G\sim 18$. Many faint sources will have low parallax signal to noise and so will only have weakly constrained distances. At these large distances, we expect that the method is using the absolute magnitude distribution combined with the measured apparent magnitude of sources to estimate the distance distribution. Oversimplifications in our absolute magnitude model therefore significantly bias the inferred halo profile. In spite of this, the thin and thick disc profiles are remarkably resilient to the halo systematics with no parameters producing significant offsets between the two fits.

As the model parameterisation used for GUMS is significantly different to our own, we cannot make direct comparisons between our parameter values and a `ground truth' input. The thick disc scale height of $670-780$ pc is broadly consistent with the scale heights used to generate the population however the GUMS sample has two profiles for different age populations which bracket our inferred value.

We conclude from this that the inferred halo profile parameters are susceptible to significant systematic uncertainties when applying our method to realistic \gaia samples likely due to an oversimplified absolute magnitude model. However the disc parameters are more reliable and we can draw information about the structure of the Milky Way disc profiles from our results when applying to the real \gaia data in \PaperII.

Even given the consistency of our disc parameter results, we cannot guarantee that we have managed to separate out the two components. Since our model is different to the one used to generate GUMS, it is likely that some thin disc stars will be contributing to the thick disc and \textit{vice versa}. This will happen to some extent with the real \gaia data too assuming that the Galactic disc can even be decomposed into discrete components which is contested \citep[e.g.][]{Bovy2012nothick}. In the scenario that the Milky Way disc does not decompose well into exponential thin and thick discs, the sum total of our disc models which describes the total tracer density may be of greater interest to the community than the individual components which contribute to it.

In \PaperII we quantify the systematic uncertainty introduced by some of the assumptions we have applied. We test the impact of Solar position offset from the Galactic mid-plane, dust extinction, magnitude uncertainty, parallax zero-point offset, shifting the turn-off absolute magnitude, Galactocentric radius dependent disc and halo distributions and an oblate stellar halo. This provides a comprehensive overview of the systematic uncertainties introduced to parameter estimates by oversimplifications in our model of stars in the Galaxy.

\section{Conclusions}

We have developed a method to fit the distribution of stars in the Milky Way using the Poisson likelihood function. Our method correctly accounts for the sample selection function and parallax measurement uncertainty.

The method is used to fit the vertical distribution of stars with $|b|>80^\circ$. For the model we use two exponential disc components and a power-law halo. The data are also simultaneously fit with a four-piece exponential absolute-magnitude distribution.

The efficacy of our method is demonstrated against a mock sample. By refitting the model parameters we demonstrate that the method produces results which are accurate to within the statistical uncertainties of the parameter posteriors.

We apply our method to the GUMS mock sample to infer the parameters of a population which are drawn from a far more complex Milky Way model. We obtain consistent fits when applying our model with and without parallax error for disc parameters but not for halo parameters. This suggests that our results for disc parameters are reliable when fit to data which doesn't exactly represent our model but that our disc parameters should be viewed with caution.

In \PaperII, we  set the machinery working on \gaia EDR3. We undertake a set of strenuous tests to quantify the systematic uncertainties in our parameter estimates due to over-simplifications in the model.

\section*{Acknowledgements}
AE thanks the Science and Technology Facilities Council of
the United Kingdom for financial support. DB thanks Magdalen College for his fellowship and the Rudolf Peierls Centre for Theoretical Physics for providing office space and travel funds. RG acknowledges financial support from the Spanish Ministry of Science and Innovation (MICINN) through the Spanish State Research Agency, under the Severo Ochoa Program 2020-2023 (CEX2019-000920-S).

This work has made use of data from the European Space Agency (ESA) mission \gaia (\url{https://www.cosmos.esa.int/gaia}), processed by the \gaia Data Processing and Analysis Consortium (DPAC, \url{https://www.cosmos.esa.int/web/gaia/dpac/consortium}). Funding for the DPAC has been provided by national institutions, in particular the institutions participating in the \gaia Multilateral Agreement. 

AE is very grateful to Eugene Vasiliev who provided valuable assistance on numerical integration methods. His comments on the paper content led to many significant improvements.

Whilst not detailed in the paper, the authors made use of the \textit{AuriGaia} mock catalogues \citep{Grand2018} which helped to motivate the model and approach taken.

The authors are grateful to the anonymous referee who suggested the use of GUMS to test our model and to Annie Robin for providing insight into the GUMS sample selection.

\section*{Data Availability}

The data underlying this article are publicly available from the European Space Agency's \gaia archive (\url{https://gea.esac.esa.int/archive/}). The EDR3 nominal scanning law is also available from the \gaia archive (\url{http://cdn.gea.esac.esa.int/Gaia/gedr3/auxiliary/commanded_scan_law/}). 

The selection function implementations used in this work are described in \citealt{CoGII} and \citealt{CoGV} and made publicly accessible through the Python package \textsc{selectionfunctions} (\url{https://github.com/gaiaverse/selectionfunctions}). The Astrometric Spread Function used to generate realistic uncertainties can be accessed through the Python package \textsc{scanninglaw} (\url{https://github.com/gaiaverse/scanninglaw}). 

The code used to fit the model and produce all figures is made publicly available as a GitHub repository (\url{https://github.com/aeverall/mwtrace.git}).



\bibliographystyle{mnras}
\bibliography{references}




\appendix

\section{Integrand limitations}
\label{app:Rintegrand}

In Section~\ref{sec:p_integrate} we stated that the integral over parallax uncertainty becomes intractable for more complex models, we will briefly justify that statement here where we will use the example of the exponential disc model to demonstrate.

The integrand including $R$-dependence is
\begin{equation}
    I\,\mathrm{d}\varpi \propto \varpi^{-4}\,\exp\left(-\frac{z}{h} - \frac{R}{L}\right)\,\exp\left(\frac{(\varpi-\varpi_i)}{2\sigma_{\varpi i}^2}\right) \,\mathrm{d}\varpi
\end{equation}
where $h$ and $L$ are the scale height and length of the disc being considered. The Jacobian for the logit transformation we applied is
\begin{equation}
    J \propto \frac{1}{(\varpi_{j+1}-\varpi)(\varpi-\varpi_j)}.
\end{equation}
Taking the gradient of $I/J$, setting to zero (as in Eq.~\ref{eq:integrand_diff}) and simplifying down, we are left with
\begin{equation}
    -\frac{4}{\varpi} - \frac{1}{h}\frac{\partial z}{\partial \varpi} - \frac{1}{L}\frac{\partial R}{\partial \varpi} - \frac{(\varpi-\varpi_i)}{\sigma_{\varpi i}^2} + \frac{1}{(\varpi_{j+1}-\varpi)} - \frac{1}{(\varpi-\varpi_j)} = 0
    \label{eq:zonly_diff}
\end{equation}
where 
\begin{equation}
z=\frac{\sin b}{\varpi}\quad \mathrm{and}\quad R^2 = R_{0}^2 + \left(\frac{\cos b}{\varpi}\right)^2 - \frac{2R_{0} \cos b\,\cos l}{\varpi}.
\label{eq:zR_def}
\end{equation}

In our application, we have assumed no R-dependence, i.e. setting $L=\infty$. We have
\begin{equation}
    \frac{\partial z}{\partial \varpi}=-\frac{\sin b}{\varpi^2}
\end{equation}
and Eq.~\ref{eq:zonly_diff} simplifies to a quintic polynomial in terms of $\varpi$. We know that at least two solutions of the quintic are outside $[\varpi_j,\varpi_{j+1}]$ since
\begin{equation}
    \frac{I}{J} \begin{cases} 
    = 0 & \mathrm{for}\,\, \varpi= \varpi_j,\varpi_{j+1} \\
    < 0 & \mathrm{for}\,\, \varpi\lesssim \varpi_j,\,\varpi\gtrsim\varpi_{j+1} \\
    = 0 & \mathrm{for}\,\, \varpi=0, \infty 
    \end{cases}
\end{equation}
so there must be a stationary point above and below the boundaries. This leaves three stationary points in the integration range corresponding to two peaks or modes. Our model is equivalent to the exponentially-decreasing square distance prior used by Section~7 of \citealt{BailerJones2015} and they also find the same two modes. However two of the roots are often either complex, or, for $\varpi_i<0$, there will be a mode with negative parallax which is outside the integration limits. Whilst we cannot guarentee that the integrand is always unimodal, Section~\ref{sec:infer} demonstrates that this does not have a measurable affect on our results.

If, however, we include $R$-dependence and have $L$ of order unity (kpc), then the integrand significantly changes. Eq.~\ref{eq:zonly_diff} now includes
\begin{equation}
\frac{\partial R}{\partial \varpi} = \frac{1}{R} \left(-\frac{\cos^2 b}{\varpi^3} + \frac{R_{0}\cos b\cos l}{\varpi^2}\right)
\end{equation}
where $R$ is given in Eq.~\ref{eq:zR_def}. Expanding this out, Eq.~\ref{eq:zonly_diff} is now an $11^\mathrm{th}$ order polynomial in $\varpi$. Again, two of the stationary points are outside the integration bounds due to the logit transformation but that leaves $9$ stationary points meaning up to $5$ modes in the integrand.

One simplification we could take which would avoid adding any more modes to the integrand is 
\begin{equation}
    R \approx R_{0} - X = R_{0} - s \cos l \cos b.
\end{equation}
This would provide a slight improvement on our previous models however also makes the model normalisation non-analytic which adds another layer of complexity. This may be an avenue worth pursuing however we consider it beyond the scope of this work.

\section{Magnitude distribution}
\label{app:mag_dist}

To derive the absolute magnitude distribution, we start from a power-law IMF with a break at $\mathcal{M}_b=0.5 M_\odot$
\begin{equation}
    f(\mathcal{M})\mathrm{d}\mathcal{M} = \begin{cases}
    \mathcal{N}_{\mathcal{M}_1}\exp\left(-\epsilon_1\right) & \mathcal{M}<\mathcal{M}_b \\
    \mathcal{N}_{\mathcal{M}_2}\exp\left(-\epsilon_2\right) & \mathcal{M}>\mathcal{M}_b.
    \end{cases}
\end{equation}
The continuity boundary condition at $\mathcal{M} = \mathcal{M}_b$ constrains $\mathcal{N}_{\mathcal{M}_1}\mathcal{M}_b^{-\epsilon_1}=\mathcal{N}_{\mathcal{M}_2}\mathcal{M}_b^{-\epsilon_2}$.

The initial-mass luminosity relation is assumed to approximately follow a set of power-law slopes
\begin{equation}
    L \propto \mathcal{M}^a
\end{equation}
such that the absolute magnitude distribution is given by
\begin{equation}
    M = \frac{-2.5 a}{\log(10)} \log(\mathcal{M}) + C
\end{equation}
where C is an unknown normalisation constant.

Fig.~\ref{fig:massmag_isochrone} shows that the magnitude-luminosity relation changes approximately around the mass break in the IMF. We introduce the magnitude boundary $M_\mathrm{MS}$ which is the absolute magnitude approximately corresponding to the mass $\mathcal{M}_b$
\begin{equation}
    M = \begin{cases}
    \frac{-2.5 a_1}{\log(10)} \log(\mathcal{M}) + C_1 & M>M_\mathrm{MS}\\
    \frac{-2.5 a_2}{\log(10)} \log(\mathcal{M}) + C_2 & M<M_\mathrm{MS}.
    \end{cases}
\end{equation}
Applying a continuity condition for the magnitude-initial mass relation at $M_\mathrm{MS}$ constrains
\begin{equation}
    C_2 = C_1 + \frac{\log(10)M_\mathrm{MS}}{2.5a_2}\left(\frac{1}{a_2} - \frac{1}{a_1}\right)
\end{equation}

We can now construct the absolute magnitude distribution.
\begin{align}
    f(M)\mathrm{d}M &= f(\mathcal{M})\mathrm{d}\mathcal{M} \\
                    &= f(M(\mathcal{M}))\mathrm{d}\left|\frac{\partial \mathcal{M}}{\partial M}\right|\mathrm{d}M\\
                    &=\mathcal{N}\begin{cases}
                    \frac{1}{a_1}\exp\left(-\alpha_1 (M-M_\mathrm{MS})\right)\mathrm{d}M & M>M_\mathrm{MS}\\
                    \frac{1}{a_2}\exp\left(-\alpha_2 (M-M_\mathrm{MS})\right)\mathrm{d}M & 
                    M<M_\mathrm{MS}\\
                    \end{cases}
\end{align}  

The model derived so far assumes a discontinuous change in the gradient of the magnitude-initial mass relation, however, Fig.~\ref{fig:massmag_isochrone} clearly shows that there's a continuous change between modes. To reflect this the model shifts between regimes across a range of apparent magnitudes. We refer to the intermediate magnitude range as the `gap' and use an extra exponential profile which connects smoothly into the lower and upper main sequences.
\begin{align}
    f(M)\mathrm{d}M &=\mathcal{N}\begin{cases}
                    \frac{1}{a_1}\exp\left(-\alpha_1 (M-M_\mathrm{MS})\right)\mathrm{d}M & M_\mathrm{MS1}<M\\
                    A_g\exp\left(-\alpha_g (M-M_\mathrm{MS})\right)\mathrm{d}M & M_\mathrm{MS2}<M<M_\mathrm{MS1}\\
                    \frac{1}{a_2}\exp\left(-\alpha_2 (M-M_\mathrm{MS})\right)\mathrm{d}M & 
                    M<M_\mathrm{MS2}\\
                    \end{cases}
\end{align}  
This introduces two boundary conditions which are continuity conditions at $M_\mathrm{MS1}$ and $M_\mathrm{MS2}$. Applying the boundary conditions fully constrains both $A_g$ and $\alpha_g$
\begin{align}
    &\alpha_g = \frac{\log\left(\frac{a_1}{a_2}\right) - \alpha_1(M_\mathrm{MS} - M_\mathrm{MS1}) + \alpha_2(M_\mathrm{MS} - M_\mathrm{MS2})}{M_\mathrm{MS1}-M_\mathrm{MS2}}\\
    &A_g = \frac{1}{a_1} \exp\left((\alpha_g-\alpha_1)(M_\mathrm{MS}-M_\mathrm{MS1})\right).
\end{align}

This gives us our main sequence distribution. However the giants follow a steeper track with a sharp drop at the turn off magnitude, $M_\mathrm{TO}$. For this, a final exponential component is included with an independent normalisation to the main sequence.
\begin{equation}
    f_G(M) \mathrm{d}M = \mathcal{N}_G \exp(-\alpha_G (M-M_\mathrm{TO})) \quad M<M_\mathrm{TO}
\end{equation}
where $\mathcal{N}_G = -\frac{1}{\alpha_G}$ to normalize the giant distribution.

Finally, putting this all together, the fraction of all sources which are dwarfs (i.e. have $M>M_\mathrm{TO}$) is parameterised by $f_D$. The full magnitude distribution is given by:
\begin{align}
    f(M)\mathrm{d}M &=\begin{cases}
    f_D\mathcal{N}_D\frac{1}{a_1}\exp\left(-\alpha_1 (M-M_\mathrm{MS})\right)\mathrm{d}M & M_\mathrm{MS1}<M<M_\mathrm{X}\\
                    f_D\mathcal{N}_DA_g\exp\left(-\alpha_g (M-M_\mathrm{MS})\right)\mathrm{d}M & M_\mathrm{MS2}<M<M_\mathrm{MS1}\\
                    f_D\mathcal{N}_D\frac{1}{a_2}\exp\left(-\alpha_2 (M-M_\mathrm{MS})\right)\mathrm{d}M & 
                    M_\mathrm{TO}<M<M_\mathrm{MS2} \\
    (1-f_D) \mathcal{N}_G \exp(-\alpha_G (M-M_\mathrm{TO}))\mathrm{d}M & M<M_\mathrm{TO}
    \end{cases}
\end{align}  

Where $\mathcal{N}_D$ is the normalisation of the full main sequence. In order to make this well normalised, an upper absolute magnitude limit, $M_\mathrm{X}$ has been placed on the lower main sequence. This also cuts the distribution off before it reaches the end of the information from isochrones. At these magnitudes there are very few visible stars and those which are in the \gaia data set will be nearby with well constrained parallaxes enabling them to be easily removed from the sample as contamination.


\bsp	
\label{lastpage}
\end{document}